\documentclass[manuscript,screen]{acmart}

\usepackage{amsmath,amssymb,amsfonts}
\usepackage{algorithmic}
\usepackage{graphicx}
\usepackage{textcomp}
\usepackage{xcolor}
\usepackage{comment}
\usepackage{balance}
\usepackage{xspace}
\usepackage{booktabs}
\usepackage{tabularray}
\usepackage[ruled, lined, linesnumbered, commentsnumbered, longend]{algorithm2e}
\usepackage{multirow,multicol}
\usepackage{tikz}
\usepackage{xcolor}
\usepackage{booktabs}
\usepackage{tabularx}
\usepackage{colortbl}
\usepackage{array}
\usepackage{enumitem}
\usepackage{tcolorbox}
\usetikzlibrary{matrix,chains,positioning,decorations.pathreplacing,arrows}
\newcommand{\sw}[1]{\textcolor{red}{{\it [Shaowei says: #1]}}}

\SetCommentSty{mycommfont}

\newcommand{\ourTool}{VFDelta\xspace}
\newcommand{\ourToolBert}{VFDelta$_{Bert}$\xspace}
\newcommand{\ourToolStar}{VFDelta$_{Star}$\xspace}

\newcommand{\vfm}{VulFixMiner\xspace}
\newcommand{\midas}{MiDas\xspace}

\newcommand{\costFive}{CostEffort@5\xspace}
\newcommand{\costTwe}{CostEffort@20\xspace}
\newcommand{\codebefore}{code\_before\xspace}
\newcommand{\codeafter}{code\_after\xspace}
\newcommand{\codechange}{diff\xspace}
\newcommand{\context}{context\xspace}

\newcommand{\EmbedConcat}{EmbedConcat\_Duo\xspace}
\newcommand{\SingleEmbed}{EmbedSubtract\_Single\xspace}
\newcommand{\CodeConcat}{CodeConcat\xspace}
\newcommand{\CodeSubtract}{CodeConcat$_{NoContext}$\xspace}
\newcommand{\RawGitDiff}{RawGitDiff\xspace}
\newcommand{\codeDelta}{Code Delta Representation\xspace}

\newcommand{\rqone}{\textit{RQ1: How effective is \ourTool compared with SOTA approaches?}\xspace}

\newcommand{\rqtwo}{\textit{RQ2: How effective is our code change representation learning?}\xspace}

\newcommand{\rqthree}{\textit{RQ3: To what extent does the context contribute to the effectiveness of \ourTool?
}\xspace}

\newcommand{\rqfour}{\textit{RQ4: What is the effectiveness of \ourTool in real-world scenario? }\xspace}

\newcommand{\rqboxc}[1]{\begin{tcolorbox}[left=1pt,right=1pt,top=1pt,bottom=1pt,colback=gray!5,colframe=gray!40!black,before skip=5pt,after skip=0pt]#1\end{tcolorbox}}

\newcommand*\circled[1]{\tikz[baseline=(char.base)]{
            \node[shape=circle,fill,inner sep=1pt,font=\footnotesize] (char) {\textcolor{white}{#1}};}}

\AtBeginDocument{%
  \providecommand\BibTeX{{%
    \normalfont B\kern-0.5em{\scshape i\kern-0.25em b}\kern-0.8em\TeX}}}
\begin{document}

\title{\ourTool: A Framework for Detecting Silent Vulnerability Fixes by Enhancing Code Change Learning}

\author{Xu Yang}
\affiliation{%
  \institution{Department of Computer Science, University of Manitoba}
  \city{Winnipeg}
  \country{Canada}}
\email{yangx4@myumanitoba.ca}

\author{Shaowei Wang}
\affiliation{%
  \institution{Department of Computer Science, University of Manitoba}
  \city{Winnipeg}
  \country{Canada}
}
\email{shaowei.wang@umanitoba.ca}

\author{Jiayuan Zhou}
\affiliation{%
  \institution{Huawei Canada}
  \city{Toronto}
  \country{Canada}
}
\email{jiayuan.zhou1@huawei.com}

\author{Xing Hu}
\affiliation{%
  \institution{Zhejiang University}
  \city{Ningbo}
  \country{China}
}
\email{xinghu@zju.edu.cn}

\renewcommand{\shortauthors}{Xu et al.}

\begin{abstract}
Vulnerability fixes in open source software (OSS) usually follow the coordinated vulnerability disclosure model and are silently fixed. This delay can expose OSS users to risks as malicious parties might exploit the software before fixes are publicly known. Therefore, it is important to identify vulnerability fixes early and automatically. Existing methods classify vulnerability fixes by learning code change representations from commits, typically by concatenating code changes, which does not effectively highlight nuanced differences. Additionally, previous approaches fine-tune code embedding models and classification models separately, which limits overall effectiveness. We propose \ourTool, a lightweight yet effective framework that embeds code before and after changes using independent models with surrounding code as context. By performing element-wise subtraction on these embeddings, we capture fine-grain changes. Our architecture allows joint training of embedding and classification models, optimizing overall performance. Experiments demonstrate that \ourTool achieves up to 0.33 F1 score and 0.63 CostEffort@5, improving over state-of-the-art methods by 77.4\% and 7.1\%, respectively. Ablation analysis confirms the importance of our code change representation in capturing small changes. We also expanded the dataset and introduced a temporal split to simulate real-world scenarios; \ourTool significantly outperforms baselines VulFixMiner and MiDas across all metrics in this setting.
\end{abstract}

\maketitle

\section{Introduction}\label{sec:intro}


Open-source software (OSS) is widely used by developers around the world. Developers must effectively manage OSS vulnerabilities, e.g., timely detection and remediation. Failure to address these vulnerabilities promptly exposes users to significant security risks, potentially leading to severe consequences~\cite{Equifax,badUnpatchedVulnerabilities}. 
To enhance OSS vulnerability management, the Coordinated Vulnerability Disclosure (CVD) process is widely adopted~\cite{MFSCVD,githubCVD,walshe2022coordinated}. This process dictates that vulnerabilities should be fixed before being publicly disclosed, enabling developers to initiate the remediation process promptly upon disclosure. Additionally, to mitigate the risk of disseminating vulnerability-related information, CVD recommends fixing vulnerabilities silently, such as refraining from including vulnerability details in commit messages. However, due to various factors like limited resources or lengthy fix-to-integration release cycles, the timing of public disclosure often deviates from the fixed date, resulting in a time gap and opening the door for attack~\cite{li2017large,zhou2021vulfixminer,imtiaz2022open}.
Due to the transparent nature of OSS, attackers can exploit this time gap to uncover fixes and derive the corresponding vulnerabilities before public disclosure. For instance, an industry report shows that unpatched vulnerabilities are directly responsible for up to 60\% of all data breaches~\cite{badUnpatchedVulnerabilities}. Therefore, developers need to detect silent fixes to start the remediation process as soon as possible.

Therefore, there is an emergence for the research community and industry to invest efforts in developing techniques to identify silent vulnerability fixes. Current approaches typically train the commit-level prediction model in two phases~\cite{zhou2021vulfixminer,midas,zhou2023colefunda}. First, they train embedding models independently on different levels of granularity by fine-tuning pre-trained models (e.g., CodeBERT~\cite{feng2020codebert}). In the second stage, they feed the joint representations produced by multiple embedding models into a new ensemble classification model, which needs to be trained as well, and predict whether the commit is to fix a vulnerability.

Existing approaches exhibit several limitations.
\circled{1} \textbf{Inefficient code change representation}. They typically represent changes as the concatenation of added code and removed code as a single sequence
~\cite{zhou2021vulfixminer,midas,zhou2023colefunda}, or as the concatenation of the resultant embeddings of added code and removed code produced by embedding models~\cite{midas}. They assume the code changes represented in the concatenated sequence or the concatenated embedding could be captured well in the new ensemble classification model. However, pre-trained language models are not trained in such a concatenated input style. Those representation approaches lack the direct induction to highlight the changes or interaction between the removed/added code, which hinders existing approaches from effectively capturing nuanced information from code changes.
\circled{2} \textbf{Missing context.} They only consider added/removed code, but do not consider the surrounding code as the context~\cite{zhou2021vulfixminer,midas,zhou2023colefunda}, which has been demonstrated to help capture the semantics of the code in various software engineering tasks~\cite{lin2022context,li2020dlfix}.
\circled{3} \textbf{Heavy training process.} They tend to be heavyweight, necessitating the training of multiple models individually in multiple stages to handle various granularity of information (e.g., lines, hunks, files, or entire commits)~\cite{zhou2021vulfixminer,midas,zhou2023colefunda}. The fine-tuning of the embedding models is separated from the final commit-level classification model's training, which under-optimizes the entire model's effectiveness.

To address the limitations mentioned above, we present \textit{\ourTool}, a lightweight yet effective framework to identify commit-level vulnerability fixes by enhancing the code change representation learning and model training process. To address the Limitations \circled{1} and \circled{2} and improve the code change representation learning, we propose \textit{\codeDelta}.
The design of \codeDelta is inspired by the principles behind word2vec~\cite{mikolov2013distributed}, which captures the semantic relationship between two words by analyzing their latent space. 
Similarly, to directly capture the difference between two pieces of code in \textit{latent} space, \codeDelta first uses two independent code embedding models to embed the code before and after the change (including the surrounding code as the context), respectively. Then the code change representation is represented as the result of the \textit{element-wise subtraction} between the two embeddings.
The resultant delta embedding is then fed into a fully connected layer (as our classification head) for file-level prediction. More specifically, in the training phase, we train a lightweight model to predict file-level vulnerability fixes (i.e., predicting whether a modified file in a commit is vulnerability-fixing or not). During the inference phase, given a commit, we aggregate the probability of all the files of the commit by averaging them as the final commit-level prediction result. \ourTool avoids ensembling multiple embedding models into the final commit-level classification model. Benefiting from the lightweight of our model architecture, the entire model (both the code embedding models and file-level classification network), could be optimized together at once during the training phase (to address Limitation \circled{3}). 


We evaluate our framework on the VFM2021 dataset~\cite{zhou2021vulfixminer}, which consists of 1,147 Java CVE patches from 146 OSS projects collected until Feb. 2021. We compare it with state-of-the-art (SOTA) baselines \vfm~\cite{zhou2021vulfixminer} and \midas~\cite{midas}.
\ourTool outperforms the baselines on all metrics and achieves an F1 of 0.33, a CostEffort@5 of 0.632, and a CostEffort@20 of 0.76, with an improvement of 77.4\%, 8.4\%, and 5.1\% over the best baseline, respectively. A higher CostEffort@L indicates that \ourTool can find more vulnerability fixes with less human inspection effort. 
Our ablation analysis shows that our code change representation learning plays a crucial role in \ourTool, and is typically effective in capturing small code changes. Furthermore, previous studies on identifying vulnerability fixes overlook evaluating their approach in the real-world setting (i.e., splitting the data in chronological order)~\cite{zhou2021vulfixminer,midas,zhou2023colefunda}, which is unrealistic~\cite{bangash2020time,jahanshahi2019does}  To fill this gap, we tested the effectiveness of \ourTool in real-world scenario. We extended the VFM dataset to December 2023 and performed temporal prediction (use data before 2021 for training and data of 2021-2023 for evaluation). \ourTool achieves a large margin of improvement over baselines, i.e., improving at least 64.5\% and 93.9\%. in terms of F1 and CostEffort@5, respectively.

In summary, this paper makes the following contributions:
\begin{itemize}
    \item We propose a lightweight yet effective framework to identify vulnerability fixes by enhancing code change representation learning and the training process.
    \item We conducted extensive experiments and show that our framework significantly outperforms all state-of-the-art baselines in identifying commit-level silent vulnerability fixes in real-world scenario.
    \item We release our enhanced temporal VFM2023 dataset to facilitate future research~\cite{anonymous_2024_10901854}.
\end{itemize}


\section{Background \& Related work}\label{sec:background}

In this section, we first present the formal definition of the problem and current solutions. Then, we introduce the background of pre-trained model leveraged in our framework. Lastly, we introduce related work.

\subsection{Silent commit-level vulnerability fix identification}
The primary objective of the task is to predict whether a commit is intended for vulnerability-fixing. The input is a commit and the output is a label either vulnerability-fixing (VF) or non-vulnerability-fixing (NVF). Commit comprises several types of information, which can be broadly categorized into three groups: commit metadata (including details such as the committer, commit date, repository, etc), commit message, and commit code changes. A commit's code change consists of alterations across multiple files, with each file's modification potentially including multiple changes at the hunk level. Models can, therefore, be designed to operate at various granularities, e.g., commit-level and file-level. 

The typical framework for training a model to identify the commit-level vulnerability fixes could be decomposed into the following phases: 1) feature extraction, 2) code change representation learning, and 3) classification training. In the feature extraction, features such as code changes and commit messages will be extracted. During code change representation learning, the extracted features will be represented into vectors (i.e., embedding). How features are extracted and how a representation is learned depends on the techniques and model design. We will introduce more details on the code change representation learning in Section~\ref{sec:embedding}. Once the representation learning is done, a binary classification model is trained to distinguish vulnerability fixes. 
Many previous works have developed mechanisms (e.g. neural network or ensemble model) to aggregate features from multiple fine-grained levels, culminating in a final prediction at the commit level. This aggregation ensures that the classifier makes a commit-level prediction based on a comprehensive view of the \codechange, facilitating accurate vulnerability-fixing commit identification.

Some early work focused on using commit messages to identify vulnerability fixes~\cite{zhou2017automated,nguyen2022hermes}. However, CVD recommends fixing vulnerabilities silently, such as refraining from including vulnerability details in commit messages, which makes it challenging to leverage the message for this task. Therefore, in this study, we focus on only using code changes to identify silent commit-level vulnerability fixes as recent work~\cite{zhou2021vulfixminer,midas}.

In this paper, we make the following definitions and use them in the following sections.
\begin{itemize}
    \item \textbf{\codechange:} The removed/added code in a file of a commit. Note that the diff we define here is distinct from the output by the Git diff algorithm, which contains three lines of surrounding code. 
    \item \textbf{\context:} The surrounding $k$ lines of code before and after the added/removed code. 
    \item \textbf{\codebefore:} The removed code plus the context before the modification.
    \item \textbf{\codeafter:} The added code plus the context after the modification.
\end{itemize}

\begin{figure}[h]
	\includegraphics[width=1.00\linewidth]{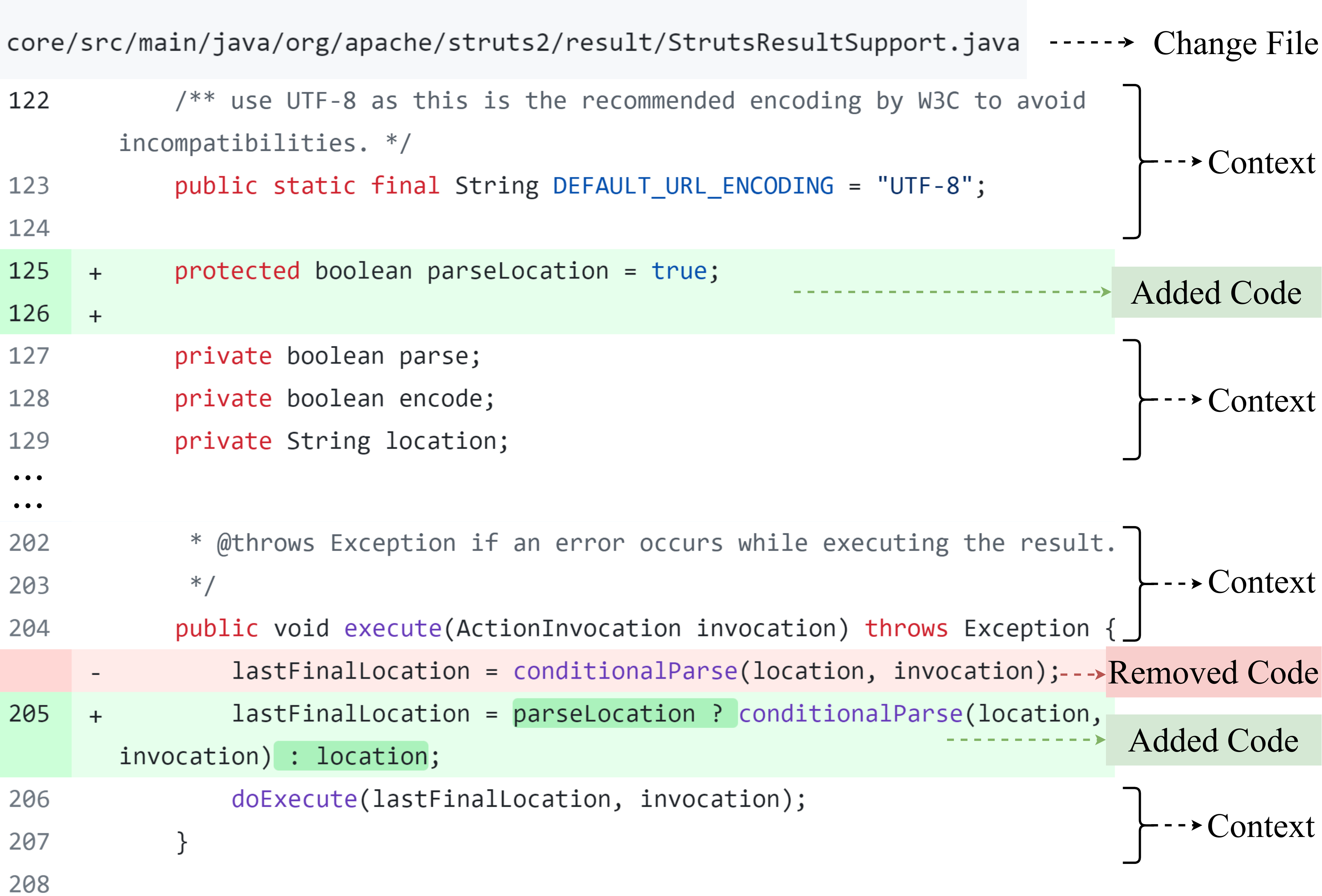}
	\caption{A sample file-level code changes for fixing CVE-2018-11776~\cite{Apache2018CVE} vulnerability.}
	\label{fig:diff_2}
\vspace{-0.1in}
\end{figure}

For instance, Figure~\ref{fig:diff_2} illustrates these definitions using an example of file-level changes aimed at fixing a CVE vulnerability. The added lines are highlighted in green, while the removed lines are highlighted in red, collectively representing the \codechange. Additionally, the surrounding lines are considered as the context. Specifically, the removed code highlighted in red, along with its context, constitutes the \codebefore, whereas the added code highlighted in green, along with its context, constitutes the \codeafter. Our study focuses on leveraging these code changes at the file level to identify commit-level vulnerability fixes.

\subsection{Code changes representation learning using pre-trained models}\label{sec:embedding}
Several pre-trained models tailored for programming languages have emerged, with CodeBERT being particularly notable~\cite{feng2020codebert}. CodeBERT is a model that builds upon the foundations laid by BERT~\cite{devlin2018bert}, further refined through pre-training on natural language (NL) - programming language (PL) pairs. Its design and training methods enable it to understand not only the relationship between natural language and code but also the semantics of the programming language. This versatility has allowed CodeBERT to demonstrate its utility across a wide range of downstream software engineering tasks~\cite{zhou2021assessing}, such as clone detection~\cite{tao2022c4,sonnekalb2022generalizability}, code search~\cite{shi2022better}, program repair~\cite{mashhadi2021applying}, and vulnerability detection~\cite{yang2023does,10.1145/3468264.3468597}. In vulnerability fix identification, pre-trained models such as CodeBERT are usually used to represent the code changes in two ways, 1) use the models to embed \codechange directly~\cite{zhou2021vulfixminer,midas}, or use models to embed removed and added code separately and then concatenate their resultant embeddings~\cite{midas}. 

\subsection{Related work in identifying vulnerability fixes}

Identifying commit-level vulnerability fixes has been one active area of research, for which different methodologies have been undertaken to improve the accuracy and efficiency of the approaches~\cite{haixiang2017learning,khoshgoftaar2015ensemble,dong2020survey, zhou2017automated, nguyen2022hermes, chen2020machine,zhou2021vulfixminer,xu2017spain,hoang2019deepjit,hoang2020cc2vec,zhou2023colefunda,midas}.
The related work can be categorized into two families based on whether they utilize commit messages or not.

The first family uses the commit message in their framework. Early work by Zhou and Sharma~\cite{zhou2017automated} proposed an approach to identify vulnerability fixes by leveraging word2vec to capture the latent information represented in commit messages. They ensemble multiple classifiers to robust the performance. 
Sabetta and Bezzi~\cite{sabetta2018practical} used both code changes and commit messages to identify the vulnerability fixes by representing both the code and message as natural language. Nguyen et al.~\cite{nguyen2022hermes} combine not only information within a commit such as its commit message, but also information from the issues associated with the commit, to boost the performance. 

However, the approaches in the first family may not work well when vulnerabilities are recommended to be fixed silently and details refrain from being included in the commit messages. To tackle this, recently developed approaches only focus on learning the patterns from code changes of commits to identify vulnerability fixes.
For example, Zhou et al.~\cite{zhou2021vulfixminer} extracted \codechange from each file in a commit and used a pre-trained language model CodeBERT to embed it. To boost the performance of CodeBERT, they fine-tuned it first and then fed the embedding of \codechange into a new classifier for the final training. 
Later, Zhou et al. introduced CoLeFunDa~\cite{zhou2023colefunda} that leverages contrastive learning to train a function change encoder from diverse data effectively. Then, they leveraged the trained function change encoder to further fine-tune for three downstream tasks and demonstrated its effectiveness on those three downstream tasks. Nguyen et al. proposed to combine different granularities of information in a commit (commit-level, file-level, hunk-level, and line-level) to capture various aspects of code changes by using CodeBERT to embed each granularity of information~\cite{midas}. However, previous approaches separate the training of the embedding model(s) and the final classification model due to the complexity of their framework, which under-optimizes the performance of the final classification model. In addition, they typically represent changes as the concatenation of added code and removed code as a single sequence, or as the concatenation of the resultant embeddings of added
code and removed code produced by the embedding model. Those representation approaches lack the direct induction to highlight the changes or interaction between the removed/added code. Therefore, to tackle those limitations, we improve the code change representation learning and enhance the training process by optimizing embedding models and the final classification model jointly.

\section{Methodology}\label{sec:method}

\begin{figure*}[t!]
	\includegraphics[width=1\textwidth]{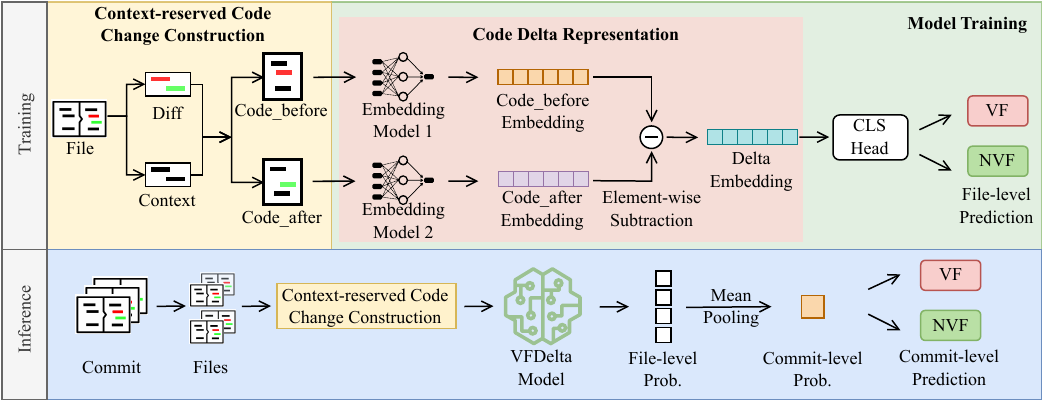}
	\caption{The workflow of \ourTool.}
	\label{fig:appraoch_pipeline}
 \vspace{-0.2in}
\end{figure*}

In this section, we introduce the methodology of \ourTool. Figure~\ref{fig:appraoch_pipeline} shows the overall framework, which consists of two phases: the file-level vulnerability-fixing classification model training and the commit-level inference.
In the training phase, we first extract file-level code changes from each commit and construct context-reserved code change information, i.e., the \textit{\codebefore} and \textit{\codeafter}.
Our model, \ourTool, is then trained to predict file-level vulnerability fixes based on the \codebefore and \codeafter. We introduce a lightweight yet effective approach (i.e., \textit{\codeDelta}) for generating file-level code change representation, directly reflecting the difference between the original code and modified code from their latent space.
The resultant representation is then fed into the final classification head for training. 
Since we train the embedding and the classifier simultaneously, we can train the entire model in one phase to optimize model performance.
During the inference phase, given a commit, we first perform file-level prediction to obtain the probability of each file, then aggregate the probability across all the files to produce the commit-level prediction result. Below, we elaborate on the details of each phase and component.
We also explain the motivation of \codeDelta in Section~\ref{sec:why_substraction}.

\subsection{File-level vulnerability-fixing classification model training}

\subsubsection{Context-reserved Code Change Construction}

In the data construction phase, we decompose a commit into file level and then extract \codebefore and \codeafter from each file by following previous studies~\cite{zhou2023colefunda,zhou2021vulfixminer}. We consider all files in a VF commit as VF and files in an NVF commit as NVF. Specifically, for each file, we first split the \codechange into two segments, i.e., code removed and code added. We then extract the surrounding code as context. Context is important for learning the difference between two pieces of code, and has been demonstrated by previous studies~\cite{lin2022context,li2020dlfix}. We extract $k$ lines of code before and after the \codechange. In this study, we empirically set $k$ to 3 as it achieves the best performance. We conducted a sensitivity analysis to evaluate the impact of different context sizes in Section~\ref{sec:rq2}.
Note that we select to use the surrounding lines of code as the context since we aim to keep our framework lightweight (e.g., avoid heavy program analysis). Nevertheless, more advanced context extraction approaches, such as program slicing, could be adapted to our framework. 

\subsubsection{\codeDelta}\label{sec:model_arch}

\codeDelta takes \codebefore and \codeafter as the input and returns the representation of their difference. Firstly, we use two individual embedding models to embed the input code, one for \codebefore and one for \codeafter. The reason we select to use two independent embedding models is that we aim to amplify and capture the nuance between \codebefore and \codeafter since we increase the tune-able parameters compared to using only one identical model for embedding (evidenced by results in Section~\ref{sec:rq2}). 
The reason we select to use two independent embedding models is that each model can specialize in capturing the characteristics of \codebefore and \codeafter without interference from the other's learning process.
Secondly, to capture the essence sometimes nuances of the modifications between \codebefore and \codeafter, we compute the delta between them through element-wise subtraction of their embeddings as shown in Figure~\ref{fig:appraoch_pipeline}. 
This design is inspired by word2vec~\cite{mikolov2013distributed}, which captures the relationship between two words using element-wise subtraction between the two embeddings. For instance, word2vec captures the relationship between gender in monarchical using subtraction, e.g.vec(King) - vec(Man) $\simeq$ vec(Queen) - vec(Women) = $\Delta$, where $\Delta$ captures the semantic relationship between the gender and royalty. Similarly, the code changes could be represented as $\Delta$ = $embedding_1(code\_before)$ - $embedding_2(code\_after)$, in which $\Delta$ captures the semantic relationship of code changes.
In addition, it is worth noting that our framework is flexible to any code embedding model. This flexibility ensures that our model can leverage the most advanced model available to better learn and understand code changes' nuances (as shown in Section~\ref{sec:rq1}).


\subsubsection{Model training}
Upon establishing the representation of code changes $\Delta$, we pass it through a feed-forward neural network leading to a classification head designed to distinguish between VF and NVF files. The entire network is trained on a dataset where each file-level change is labeled as either VF or NVF, thus learning to identify potential vulnerability fixes based on the characteristics of the code modifications. Note that different from previous approaches\cite{zhou2021vulfixminer,zhou2023colefunda,midas}, in which the model has to be trained in multiple phases (e.g., fine-tune the embedding models and train the final commit-level classification model), we fine-tune embedding models and final classification model in one phase.

\subsection{Commit-level Inference}
Once trained, the model's inference process is geared towards predicting whether an unseen commit is VF or not. This inference is not merely a file-level analysis but extends to a comprehensive commit-level assessment. To achieve this, we aggregate the predictions for all files within a commit by calculating their mean prediction probability as $\text{ Prediction}(Commit) = \frac{1}{N} \sum_{i=1}^{N} \text{Prediction}(File\_i)$, as $N$ denotes the number of files in a commit. We consider a commit with a probability larger than 0.5 as VF, otherwise NVF. Although, in theory, we could add and train an extra layer or model to aggregate the file-level prediction probability and learn the weights during the training phase, we decided to use a simple mean pooling since we found it is effective for predicting commit-level vulnerability fix. 

\subsection{Why element-wise subtraction?}
\label{sec:why_substraction}
Apart from the above-mentioned intuitive example, we would like to elaborate on why we select element-wise subtraction, as previous work~\cite{hoang2020cc2vec,tian2020evaluating,liu2023ccrep} have also used element-wise subtraction as one part of their code change representation, although they only used one identical embedding model to embed removed/added separately, which is different from our architecture. However, none of them explain why. We argue that element-wise subtraction is a special case of concatenation with inductive bias~\cite{mou-etal-2016-natural}, and this induction helps with code change representation learning. 
Consider two vector representations $e_1$ and $e_2$, corresponding to the embedding of \codeafter and \codebefore, $e_1 - e_2$ represents the element-wise subtraction of the two vectors, the concatenation of two vectors side by side can be represented as $[e_1,e_2]^{\top}$. Specifically, if we apply the transformation $W[e_1,e_2]^{\top}$, with $W$ being the associated weight matrix $[w_1,w_2]$ in the model. The operation of element-wise subtraction $e_1-e_2$ can be obtain through the transformation $W_0(e_1-e_2) = [w_0,-w_0][e_1,e_2]^{\top}$. Here, $W_0$ represents the weights associated with the element-wise subtraction and is a special case of $[w_1,w_2]$. This perspective shows element-wise subtraction as a special form of concatenation facilitated by a subsequent linear transformation. We use element-wise subtraction by explicitly focusing on differences between two vectors, and it introduces a strong induction that aids the learning process. Models leveraging this heuristic can more effectively learn to discern patterns of code changes and converge faster (see our analysis in Section~\ref{sec:rq2}).




\section{Experimental Setting}\label{sec:experimentalsetting}
In this section, we present research questions (RQs), datasets, evaluation metrics, our analysis approach for RQs, and implementation details.

\begin{table}[]
\caption{Statisics of our studied datasets.}
\begin{tabularx}{\textwidth}{lXXXXXXXXX}
\toprule
 & \multicolumn{9}{c}{\textbf{VFM2021   Dataset (Cross-Project Setting)}} \\ \midrule
 & \multicolumn{3}{c}{\textbf{Training Set}} & \multicolumn{3}{c}{\textbf{Validation Set}} & \multicolumn{3}{c}{\textbf{Testing Set}} \\ \cmidrule(lr){2-10} 
 & \textbf{\#V.F.} & \textbf{\#N.V.F} & \textbf{\#Project} & \textbf{\#V.F.} & \textbf{\#N.V.F} & \textbf{\#Project} & \textbf{\#V.F.} & \textbf{\#N.V.F} & \textbf{\#Project} \\
\textbf{File} & 1889 & 72635 & 117 & 163 & 7564 & 115 & 596 & 207979 & 29 \\
\textbf{Commit} & 796 & 30120 & 117 & 135 & 6304 & 115 & 242 & 87457 & 29 \\ \bottomrule
 &  &  &  &  &  &  &  &  &  \\ \toprule
 & \multicolumn{9}{c}{\textbf{VFM2023 Dataset (Temproal Setting)}} \\ \midrule
 & \multicolumn{3}{c}{\textbf{Training Set}} & \multicolumn{3}{c}{\textbf{Validation Set}} & \multicolumn{3}{c}{\textbf{Testing Set}} \\
 & \multicolumn{3}{c}{(2012/06 - 2019/12)} & \multicolumn{3}{c}{(2019/12 - 2021/01)} & \multicolumn{3}{c}{(2021/02 - 2023/12)} \\ \cmidrule(lr){2-10} 
 & \textbf{\#V.F.} & \textbf{\#N.V.F} & \textbf{\#Project} & \textbf{\#V.F.} & \textbf{\#N.V.F} & \textbf{\#Project} & \textbf{\#V.F.} & \textbf{\#N.V.F} & \textbf{\#Project}\\
\textbf{File} & 2407 & 93908 & 145 & 241 & 8119 & 127 & 892 & 87138 & 9 \\
\textbf{Commit} & 980 & 37783 & 145 & 109 & 3164 & 127 & 285 & 31374 & 9 \\ \bottomrule
\end{tabularx}
\label{tab:dataset}%
\end{table}



\subsection{Research Questions}
We evaluate \ourTool in different aspects to answer the following research questions.

\begin{itemize}
    \item\rqone
    \hfill
    \item \rqtwo
    \hfill
    \item \rqthree
    \hfill
    \item \rqfour
    
\end{itemize}

In RQ1, we aim to evaluate the performance of \ourTool by comparing it to current SOTA approaches. We propose a simple yet effective approach to learning the representation of code changes. Therefore, in RQ2, we aim to understand the contribution of the representation learning approach by comparing it to various variants. In RQ3, we aim to investigate the impact of context on \ourTool. Lastly, in RQ4, we aim to conduct a rigorous evaluation of \ourTool in a challenging real-world scenario, i.e., using the historical data for training and future data for evaluation.

\subsection{Dataset preparation}
\subsubsection{VFM2021 dataset}
To evaluate the effectiveness of \ourTool in identifying silent vulnerability fixes, we used the Java dataset proposed by recent work VulFixMiner~\cite{zhou2021vulfixminer}, short as VFM2021. VFM2021 dataset includes fixes for vulnerabilities disclosed by February 07, 2021. It contains 1,474 vulnerability-fixing commits from 150 Java repositories and 474,555 non-vulnerability-fixing commits from the same repositories. A key feature of the VFM2021 dataset is its project-level separation, ensuring no repository appears in both the training/validation/testing sets. Furthermore, to mitigate the challenge of label imbalance, the non-vulnerability fixing (NVF) commits in the training and validation sets were down-sampled. We evaluate RQ1 - RQ3 on VFM2021.

The origin VFM2021 dataset itself only contains added code and removed code (i.e., \codechange), while our approach requires the surrounding code as the context to achieve better performance. Therefore, we updated the VFM2021 to include more surrounding context for each piece of code change. To do so, we cloned the repositories afresh and gathered all commit information via Pydriller\footnote{https://pydriller.readthedocs.io/en/latest/intro.html}, which is a framework that helps developers for mining software repositories. Any commits untraceable by Pydriller or repositories that are no longer accessible on GitHub were omitted. In total, we ended with 1,172 commits. Table~\ref{tab:dataset} shows the statistics of the updated VFM2021.

\subsubsection{Enhanced temporal VFM2023 dataset}

To answer RQ4, we need to construct a dataset that is split chronologically. To do so, we extend the original VFM2021 dataset to December 2023. First, we re-divided the entire VFM2021 data set according to time. We split VFM2021 into a training set with the first 90\% VF commits and a validation set with the last 10\% VF commits.
To form the test set, we looked for Vulnerability Fixes (VFs) after 2021. First, we collected all commits between 2021 to 2023 from the VFM2021's Open Source Software (OSS) repositories. We then cross-referenced these commits with records in the National Vulnerability Database (NVD) and Snyk Vulnerability Database from 2021 to 2023 to identify VF commits by following previous study~\cite{zhou2021vulfixminer}.  From this process, we found 527 VFs in 52 repositories. To ensure dataset quality and proper evaluation of the model on adequately represented OSS repositories, we removed repositories with less than 10 VFs in either the training or testing set. Ultimately, the refined dataset comprised 9 repositories with 285 VFs and 31,374 Non-Vulnerable Fixes (NVFs). The detailed statistics of this dataset are presented in Table~\ref{tab:dataset}.

\subsection{Evaluation metrics}
We evaluate the effectiveness of \ourTool from two perspectives: classification effectiveness and effort-awareness. 
We use F1-Score(short as F1), precision, and recall as the classification effectiveness metric. Those metrics are widely used in classification tasks~\cite{yang2023does,zhou2020studying}.
Second, to measure the effort-awareness of our approach, we select CostEffort@L by following previous studies~\cite{midas,zhou2021vulfixminer,zhou2023colefunda}. Given the fact that vulnerability fixes are rare among the code commits (i.e., the extremely imbalanced dataset scenario), the goal of \ourTool is to help users reduce the efforts in identifying vulnerability fixes from an enormous scope of code commits. CostEffort@L is the proportion of inspected vulnerability fixes among all the actual vulnerability fixes when $L\%$ lines of source code in all commits are inspected and is computed as $\frac{n}{N}$, where $n$ is the correctly identified vulnerability fixes, $N$ denotes for the total vulnerability fixes, and L accounts for $L\%$ of total LOC of commits.

\subsection{Code embedding models}\label{sec:model}
To embed code, we select two widely used pre-trained models CodeBERT~\cite{feng2020codebert} and StarEncoder~\cite{li2023starcoder} that are trained on source code and designed for source code-related tasks. We select CodeBERT since it is also used in VulFixMiner~\cite{zhou2021vulfixminer} and MiDas~\cite{midas} as the model for code embedding so that we can compare with those baselines fairly. We select StarEncoder as it is a more recent and advanced model. However, both models have the same model parameters - 125M. We intend to investigate whether using a mode advanced more could improve the effectiveness of \ourTool.

\subsection{Implementation details}
We begin by downloading the official pre-trained model checkpoint for all evaluated models from HuggingFace. Subsequently, the PyTorch and Transformers packages are employed to perform all the experiments. We use 5e-05 as the training rate and AdamW as the optimizer, and we train the model for 10 epochs with a batch size of 128. We use bfloat16 (brain floating point) mixed precision and all experiments are done in Python 3.10. Experiments were conducted on a Linux server equipped with four Nvidia RTX 3090 GPUs, a 24-core CPU, and 128 GB of RAM.

\subsection{Approach for RQs}

\subsubsection{Approach of RQ1}
To demonstrate the effectiveness of \ourTool, we compare it with three recent SOTA baselines: VulFixMiner~\cite{zhou2021vulfixminer}, CoLeFunDa~\cite{zhou2023colefunda} and MiDas~\cite{midas} on the VFM2021 datasets in terms of all evaluation metrics. We obtain these models from the original authors and use them according to guidance from the authors and or replication packages. As we discussed in Section~\ref{sec:model}, we use two code embedding models, thereby we refer to the version of implementation that employs CodeBERT as to \ourToolBert and the version that uses StarEncoder as to \ourToolStar.

\subsubsection{Approach of RQ2}

\begin{figure}
    \centering
    \includegraphics[width=1\linewidth]{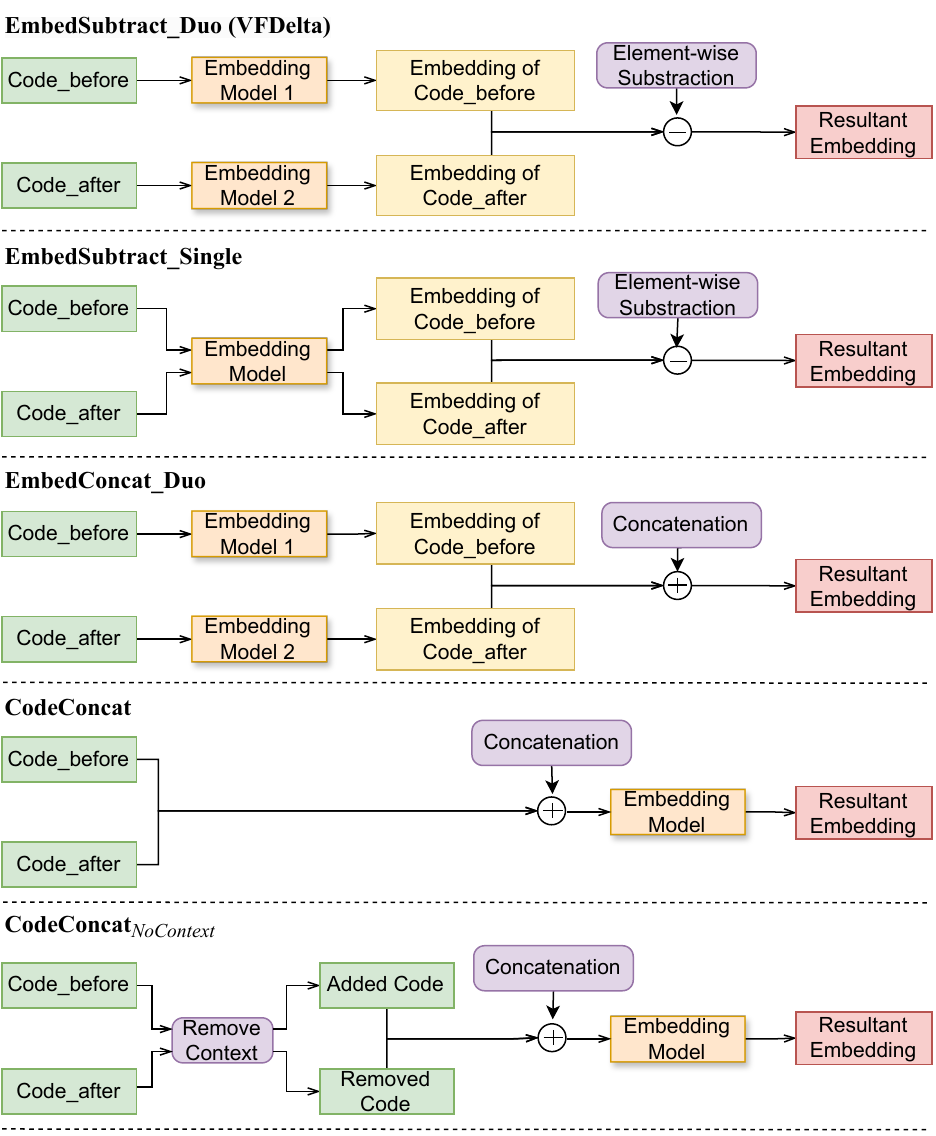}
    \caption{Comparison of different code change representation learning approaches.}
    \label{fig:rq2_pipeline}
    \vspace{-0.1in}
\end{figure}


In this RQ, we aim to evaluate the contribution of \codeDelta to \ourTool. More specifically, we first aim to elucidate the contributions of 1) using two independent embedding models for embedding \codebefore and \codeafter, and 2) adapting element-wise subtraction for embedded representation. Hence, we compare our approach \ourTool with the following two variants with different code change representation learning designs to understand the contribution of the two designs. We elaborate on the details of each variant below.


\noindent\textbf{\SingleEmbed} In this variant, we retain the element-wise subtraction of embeddings but process both \codebefore and \codeafter through a single embedding model. A similar design was used by CCRep~\cite{liu2023ccrep} to capture code change representation. This experiment aims to quantify the impact of using separate embedding models for each piece of code, allowing us to understand if the nuances and differences between \codebefore and \codeafter can be captured and represented more effectively using two independent embedding models compared with one. 

\noindent\textbf{\EmbedConcat} In this variant, we replace the element-wise subtraction operation with concatenation, in which we concatenate the outputs of the two embedding models. This variant assesses the importance of element-wise subtraction in capturing the relational dynamics between code\_before and code\_after, providing insights into how different embedding combination methods affect model performance. 

Second, we are interested in investigating the impact of performing operations (i.e., concatenation and subtraction) on the raw code and embedding (before and after the embedding). 

\noindent\textbf{\CodeConcat} In this variant, raw \codebefore and \codeafter are concatenated with a separate token (i.e., [\codebefore, sep, \codeafter]) before being fed into the embedding model.

\noindent\textbf{\CodeSubtract} In this variant, due to the subtraction on the raw code, the context of  \codebefore and \codeafter are discarded. Therefore, only the changed code (i.e., [removed code, sep, added code]) will be retained. This design was adopted by previous studies MiDas~\cite{midas} and Vulfixminer~\cite{zhou2021vulfixminer}. This variant allows us to compare with the code change representation of previous approaches directly. 

\noindent\textbf{\RawGitDiff} In this variant, we do not conduct any pre-processing to the code diff generated by the Git Diff algorithm, the input is represented as a sequence [3 lines of context before, added code, removed code, 3 lines of context after]. We use it directly as input to the embedding model. This design simulates the most simple approaches to represent code change.

Figure~\ref{fig:rq2_pipeline} provides the comparison of our approach with the studied variants. Note that since \ourTool employs two individual embedding models and subtraction, therefore we name it EmbedSubstract\_Duo in the Figure.

In addition, to better understand the capacity of \ourTool, particularly in capturing small code changes in a commit. We measure the effectiveness of each studied representation approach on commits with different sizes of code changes (i.e., number of added/removed lines of code). We group the commits into the following buckets (1,20], (20, 40], (40,60], (60, 80], (80, 100], and (100,$\infty$). 

In this RQ, we focus on CodeBERT as the base model, and keep the same setting as RQ1.

\subsubsection{Approach of RQ3}
In this RQ, we aim to investigate the impact of the window size of context on the effectiveness of \ourTool. We set context size to 0, 1, 3, 5, 7, 9. We re-train our models with different context sizes and evaluate their performance accordingly. To accommodate as much context as possible within the token limit of the pre-trained model, we evaluate \ourTool with StarEncoder, which supports up to 1024 tokens input. We keep all other experimental settings the same as RQ1.

\subsubsection{Approach of RQ4}

In the previous RQs, we assessed the performance of \ourTool utilizing the VFM2021 dataset. This dataset is organized based on open-source software (OSS) projects (cross-project setting), which may not adequately simulate real-world scenarios where the emergence of vulnerabilities (VFs) is inherently unpredictable. To address this limitation and more accurately reflect real-world application scenarios, this RQ aims to examine the efficacy of \ourTool in navigating the complexities of temporal data. Specifically, this RQ uses the VFM2023 temporal dataset, which was crafted to represent a real-world challenge where the model is trained on historical data to forecast future vulnerability fixes (VF). We compare \ourTool with the baselines (i.e., \vfm and \midas) to provide a comprehensive understanding of their performance in predicting emergent vulnerabilities within the evolving landscape of OSS projects. Similar to RQ1, we compare both the \ourToolBert and \ourToolStar with other baseline models.

\section{Results}\label{sec:results}

\begin{table}[t]
\vspace{-0.1in}
\centering
\caption{Comparison of \ourTool with baselines on VFM\_2021 dataset. We highlight the cells with the best performance in bold.}
\begin{tblr}{
  width = \linewidth,
  colspec = {X[0.8,l] X[0.3,c] X[0.5,c] X[0.4,c] X[1,c] X[1,c]},
  hline{1,7} = {-}{0.08em},
}
\textbf{Method}         & \textbf{F1}   & \textbf{Precision} & \textbf{Recall} & \textbf{\costFive} &  \textbf{\costTwe} \\ \hline
\vfm    & 0.186 & 0.113     & \textbf{0.540}   & 0.583         & 0.619          \\

\midas          & 0.120 & 0.072     & 0.240   & 0.484          & 0.723         \\ 
CoLeFunDa    & 0.062 & 0.033     & 0.533   & 0.590         & 0.721          \\\hline
\ourToolBert & 0.316 & 0.262     & 0.360  & 0.616         & 0.744          \\
\ourToolStar & \textbf{0.330}  & \textbf{0.285}     & 0.392  & \textbf{0.632}         & \textbf{0.760}     \\
\label{tab:rq1}
\end{tblr}
\vspace{-0.2in}
\end{table}
\subsection{Results of RQ1}\label{sec:rq1}

\textbf{\ourToolBert improves the SOTA baselines at least by 69.9\%, 5.7\%, and 2.9\% in terms of F1, \costFive, and \costTwe, respectively.}
Table~\ref{tab:rq1} presents the results of our approaches and current state-of-the-art approaches, \vfm and \midas.
First, for a fair comparison, we compare \ourToolBert with \midas, CoLeFunda and \vfm, which all use codeBert as their embedding model, \ourToolBert achieves an F1 of 0.316, and outperforms \vfm (0.186), CoLeFunDa (0.062) and \midas (0.120) with an improvement of 69.9\% , 409.7\% and 163.3\%, respectively. This better F1 achieved by \ourToolBert is attributed to the better precision, although we have a worse recall compared with \vfm, which suggests our approach can reduce false positives significantly. 
When examining cost-efficiency metrics, \ourToolBert improves \costFive to 0.616 compared to \vfm's 0.583, CoLeFunDa's 0.590 and \midas's 0.484, achieving an improvement of 5.7\%, 4.4\% and 26.5\%, respectively. Similarly, \ourToolBert achieves an improvement of 20.2\%, 3.2\% and 2.9\% over \vfm, CoLeFunDa and \midas, respectively, in terms of \costTwe.

Note that \ourTool achieves much better precision while sacrificing some recall, typically compared with \vfm. Higher precision indicates the reduction of false positives. Note that in real-world scenarios, reducing false positives is important as developers may lose confidence in using vulnerability detection tools for auditing if the tools produce many false positives~\cite{cheirdari2018analyzing,jovanovic2006pixy}.

\textbf{\ourToolStar achieves the best performance among all baselines including \ourToolBert, benefiting from its more advanced embedding model.}
\ourToolStar, which uses the StarEncoder as the embedding model achieves better performance than \ourToolBert. \ourToolStar improve \ourToolBert's F1 \costFive, and \costTwe from 0.316, 0.616, and 0.744 to 0.33, 0.632, and 0.76, typically achieve a 4.4\% improvement in terms of F1. If we compare \ourToolStar with \midas, CoLeFunDa and \vfm, \ourTool can improve the baselines at least by 77.4\%, 7.1\%, and 5.1\% in terms of F1, \costFive, and \costTwe. \ourToolStar demonstrates potential benefits from utilizing more advanced embedding models. The results indicate that while foundational models like CodeBert provide a solid base, exploring and incorporating cutting-edge models can lead to further notable performance improvements.

\rqboxc{\ourTool improves SOTA baselines significantly. For instance, \ourToolStar achieves the best performance on all metrics, and achieves an F1 of 0.33, a \costFive of 0.632, and a \costTwe of 0.76, with the improvement of 77.4\%, 7.1\%, and 5.1\% over the best baselines.}

\subsection{Results of RQ2}\label{sec:rq2}

\begin{table*}[t]
\centering
\caption{Comparison of \ourTool and studied variants on VFM2021.}
\renewcommand{\arraystretch}{1.1} 
\resizebox{0.95\textwidth}{!}{
\begin{tabular}{llllll}
\toprule
\textbf{Variant} & \textbf{F1} & \textbf{Precision} & \textbf{Recall} & \textbf{\costFive} & \textbf{\costTwe} \\ \midrule
\EmbedConcat & 0.226 (28.5\% $\downarrow$) & 0.155(45.0\% $\downarrow$) & 0.410(13.9\% $\uparrow$) & 0.591(4.1\% $\downarrow$) & 0.682(8.3\% $\downarrow$) \\
\SingleEmbed & 0.219 (30.7\% $\downarrow$)& 0.149(47.2\% $\downarrow$) & 0.409(13.6\% $\uparrow$) & 0.562(8.8\% $\downarrow$) & 0.661(11.2\% $\downarrow$) \\
\CodeConcat & 0.221 (30.1\% $\downarrow$)& 0.149(47.2\% $\downarrow$) & \textbf{0.430}(19.4\% $\uparrow$) & 0.566(8.1\% $\downarrow$) & 0.690(7.3\% $\downarrow$) \\
\CodeSubtract & 0.236 (25.3\% $\downarrow$)& 0.166(41.1\% $\downarrow$) & 0.413(14.7\% $\uparrow$) & 0.566(8.1\% $\downarrow$) & 0.678(8.9\% $\downarrow$) \\
\RawGitDiff & 0.216 (31.6\% $\downarrow$)& 0.144(48.9\% $\downarrow$) & \textbf{0.430}(19.4\% $\uparrow$) & 0.570(7.5\% $\downarrow$) & 0.711(4.4\% $\downarrow$) \\ \midrule
EmbedSubstract\_Duo (\ourTool) & \textbf{0.316} & \textbf{0.282} & 0.360 & \textbf{0.616} & \textbf{0.744} \\
\bottomrule 
\end{tabular}
}
\label{tab:rq2}%
\vspace{0.1in}
\end{table*}

\begin{figure}
    \centering
    \begin{tikzpicture}
        \node[anchor=south west,inner sep=0] (image) at (0,0) {\includegraphics[width=0.8\linewidth]{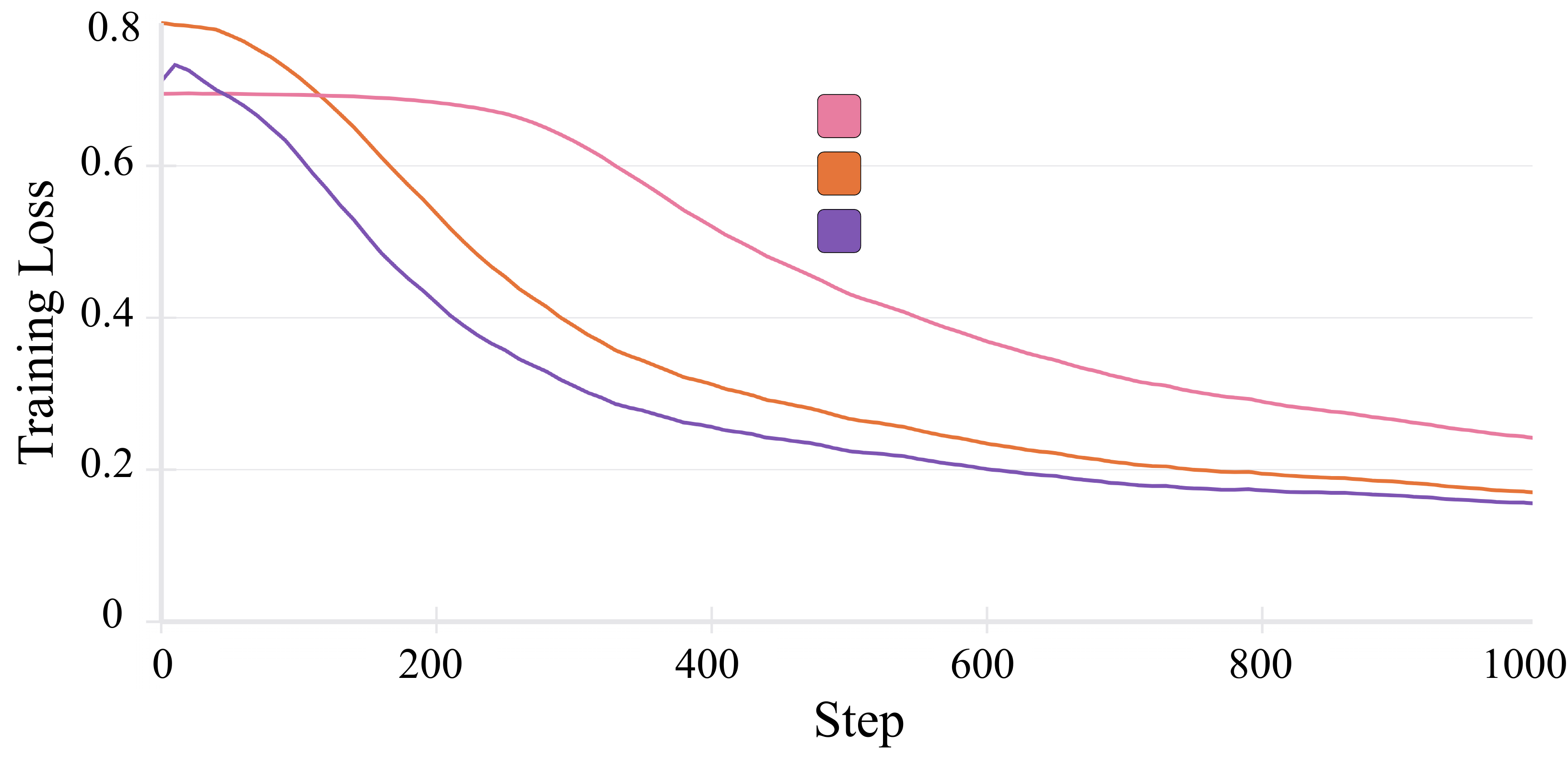}};
        \begin{scope}[x={(image.south east)},y={(image.north west)}]
            \node[font=\footnotesize,rounded corners,anchor=west] at (0.55,0.84) {\SingleEmbed};
            \node[font=\footnotesize,rounded corners,anchor=west] at (0.55,0.77) {\EmbedConcat};
            \node[font=\footnotesize,rounded corners,anchor=west] at (0.55,0.69) {EmbedSubstract\_Duo (\ourTool)};
        \end{scope}
    \end{tikzpicture}
    \caption{Comparison of training loss between \ourTool, \SingleEmbed and \EmbedConcat.}
    \label{fig:loss_diff}
\end{figure}

\textbf{Using independent embedding models for code\_before and code\_after and element-wise subtraction to capture the difference between two embeddings as a combination plays a critical role in our approach.} Table~\ref{tab:rq2} summarizes the results of each variant. Compared to \ourTool, \EmbedConcat, which concatenates embedding outputs instead of element-wise subtraction, has a 28.5\% decrease in F1 score, 4.1\% drop in \costFive, and 8.3\% drop in \costTwe. This result demonstrates using the subtraction operation to capture the differences between two embeddings is more effective than concatenation. As discussed in Section~\ref{sec:method}, subtraction is the special case of concatenation, using element-wise subtraction in the model provides a direct induction to the model to capture the difference in such a way and facilitate the learning process. This is evidenced in Figure~\ref{fig:loss_diff}, where we can observe that the training loss of \ourTool drops faster than \EmbedConcat and \SingleEmbed.
\SingleEmbed processes both pieces of code through the identical embedding model but retains element-wise subtraction. This variant results in a 30.7\% decrease in F1, 8.8\% drop in \costFive, and 11.2\% in \costTwe compared to \ourTool, which demonstrates the importance of using two independent embedding models for code representations. More importantly, as we can see, only using one strategy does not improve the performance much compared with other variants. Two individual embedding models and element-wise subtraction need to be used as a combination to optimize their effectiveness.


When comparing the two families of variants that apply operations (e.g., subtraction and concatenation) before and after embedding models, we find that they have similar performance in all metrics. \CodeConcat, \CodeSubtract, and \RawGitDiff share similar performance as \EmbedConcat and \SingleEmbed, which indicates that capturing differences in raw code and embeddings have similar effectiveness for the VF identification without a strong induction. Also, the performance of \CodeConcat and \CodeSubtract shows that providing a context within the raw code level is not beneficial.

\begin{figure}[t]
	\includegraphics[width=1.0\linewidth,trim={0.2cm 0 1.4cm 0.1cm},clip]{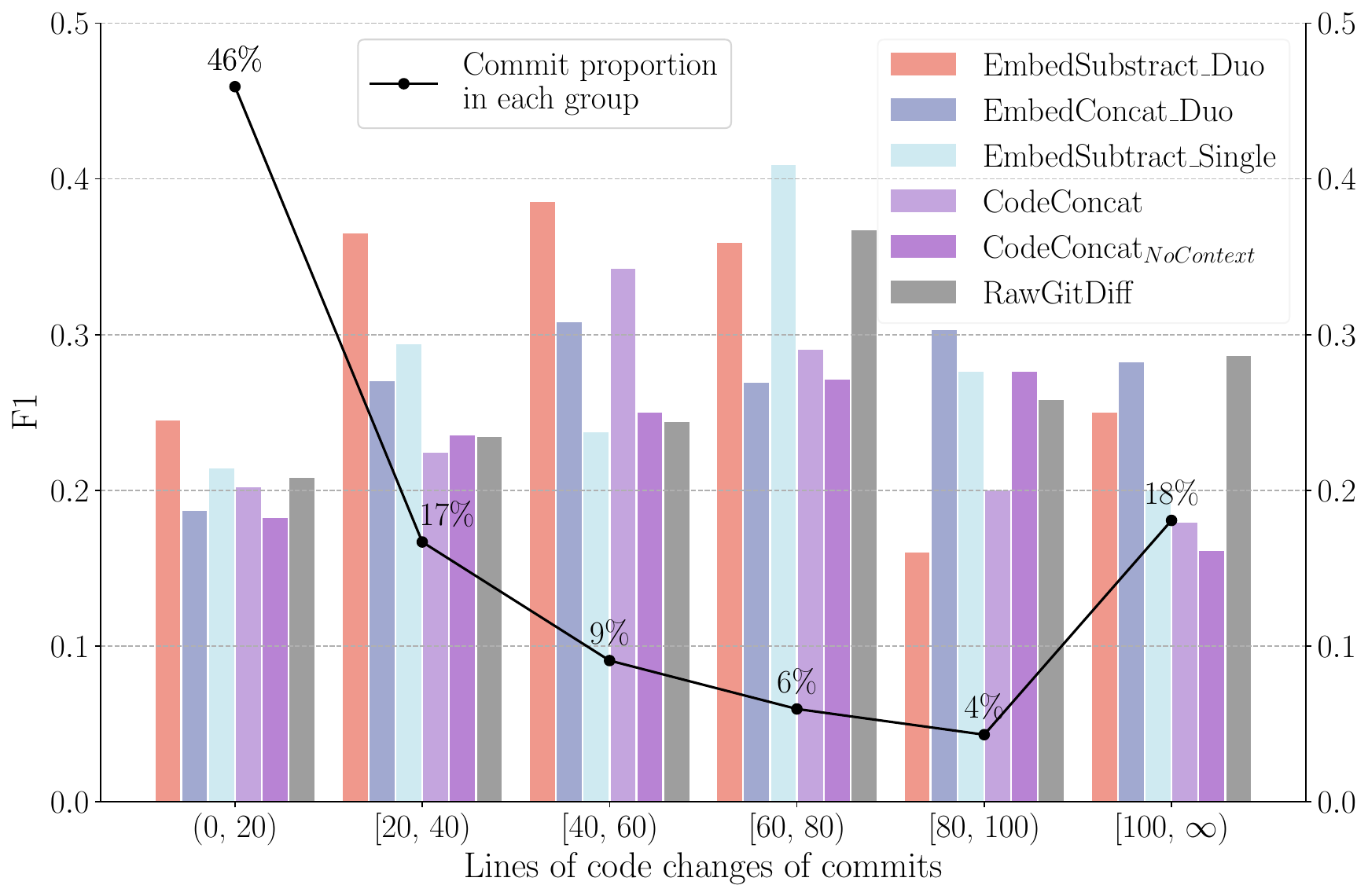}
 
	\caption{The barplot denotes the effectiveness of different approaches on the groups of commits with different change sizes in terms of F1. The black line denotes the proportion of each group of commits. For instance, 46\% indicates 46\% of the commits have a code change size between 0 and 20.}
	\label{fig:perf_loc}
\vspace{-0.15in}
\end{figure}


\textbf{Our approach of learning code change representation is more effective in capturing small changes than other studied variants and achieves better performance in identifying VF with small changes.}
Figure~\ref{fig:perf_loc} presents the performance (F1) of the studied variants in the groups of commits with different sizes of code changes, and the data distribution of each group. As observed, \ourTool outperforms other approaches when the code change is small (i.e., LOC is 0 -60). It is worth noting that the majority of the commits (72\%) fall into this range, and our approach performs better for the majority of commits. This analysis provides evidence that our design is more effective in capturing small code changes and leads to better performance in identifying VFs with small changes, typically when the changed LOC is less than 60.

\textbf{Compared to other variants, our design improve precision by a large margin and reduces false positives.} 
As we can observe, \ourTool achieves the best F1, precision, \costFive, and \costTwe, which indicates that \ourTool introduces fewer false positives. One explanation is that using two independent embedding models enables our approach to better capture the nuances between VF and NVF. This is evidenced by the best performance of \EmbedConcat among the variants (excluding \ourToolBert) typically for the small changes as shown in Figure~\ref{fig:perf_loc}, which is the only one using two embedding models to represent code\_before and code\_after. 


\rqboxc{\codeDelta plays a critical role in our approach and is typically effective in capturing small code changes.}

\subsection{Results of RQ3}\label{sec:rq3}

\begin{table}[t]
  \centering
  \caption{Performance of \ourToolStar under different context windows on VFM\_2021.}
  \begin{tabularx}{\textwidth}{>{\centering\arraybackslash}X>{\centering\arraybackslash}X>{\centering\arraybackslash}X>{\centering\arraybackslash}X>{\centering\arraybackslash}X>{\centering\arraybackslash}X}
    \toprule
    \textbf{Window size} & \textbf{F1} & \textbf{Precision} & \textbf{Recall} & \textbf{\costFive} & \textbf{\costTwe} \\
    \midrule
    0     & 0.280  & 0.214 & \textbf{0.405} & 0.603 & 0.727 \\
    1     & 0.295 & 0.254 & 0.351 & 0.541 & 0.752 \\
    3     & \textbf{0.330} & \textbf{0.285} & 0.392 & \textbf{0.632} & \textbf{0.760} \\
    5     & 0.293 & 0.243 & 0.368 & 0.558 & 0.744 \\
    7     & 0.268 & 0.210  & 0.368 & 0.579 & 0.748 \\
    9     & 0.264 & 0.206 & 0.368 & 0.545 & 0.744 \\
    \bottomrule
  \end{tabularx}
  \label{tab:rq3}
\end{table}

\textbf{\ourTool achieves the optimal performance with a 3-line context window.} Table~\ref{tab:rq3} illustrates the performance of our approach using varying numbers of surrounding code lines as context.
We note that the effectiveness of \ourTool, indicated by F1 score, precision, recall, and cost-effort metrics, fluctuates with different window sizes. Remarkably, the model achieves its peak performance with a 3-line context window. Subsequently, the performance of \ourTool diminishes as the number of lines increases. This trend suggests that inadequate context may hinder accurate vulnerability detection, whereas excessive context might introduce noise and divert the model's attention from pertinent changes.
This observation underscores the critical role of context in refining the model's capability to identify vulnerability fixes precisely.

It is worth noting that even without context, \ourTool outperforms the baselines (i.e., \midas and \vfm) in terms of F1, \costFive, and \costTwe, which indicates the superiority of our framework.


\rqboxc{Too little or much context hinders the performance. \ourTool achieves the optimal performance with a 3-line context window.}

\subsection{Results of RQ4}\label{sec:rq4}

\begin{table}[t]
\centering
\caption{Comparison of \ourTool and baselines on VFM\_2023 dataset.}
\begin{tblr}{
  width = \linewidth,
  colspec = {X[0.8,l] X[0.3,c] X[0.5,c] X[0.4,c] X[1,c] X[1,c]},
  hline{1,7} = {-}{0.08em},
}
\textbf{Method}         & \textbf{F1}   & \textbf{Precision} & \textbf{Recall} & \textbf{\costFive} &  \textbf{\costTwe} \\ \hline
\vfm    & 0.053 & 0.058     & 0.049   & 0.133         & 0.228          \\
\midas          & 0.121 & 0.103     & 0.147   & 0.228          & 0.589         \\ 
CoLeFunDa          & 0.080 & 0.052     & 0.168   & 0.218          & 0.575         \\ \hline
\ourToolBert & 0.158 & 0.162     & 0.154  & 0.351         & 0.540          \\
\ourToolStar & \textbf{0.199}  & \textbf{0.165}     & \textbf{0.256}  & \textbf{0.442}         & \textbf{0.618} 
\label{tab:rq4}
\end{tblr}
\vspace{-0.15in}
\end{table}

\textbf{\ourTool significantly outperforms the SOTA baselines \vfm and \midas across all metrics in this temporal dataset setting.}. For instance, \ourToolStar achieves an F1 of 0.199, demonstrating a notably higher prediction performance than \midas (0.121), CoLeFunDa (0.080) and \vfm (0.053), with an improvement of 64.5\%, 148.8\% and 275.5\%, respectively. For \costFive and \costTwe, we observe a similar trend. \ourToolStar improves \midas which outperforms \vfm and CoLeFunDa, from 0.228 to 0.442 for \costFive and 0.589 to 0.613 for \costTwe, with an improvement of 93.9\% and 4.08\%.
Effort-awareness metrics are particularly important in the real-world scenario under the constraints of limited resources or inspection capacity. Our observations demonstrate the efficiency of \ourTool in real-world scenarios.

\begin{figure}[h]
	\includegraphics[width=0.8\linewidth]{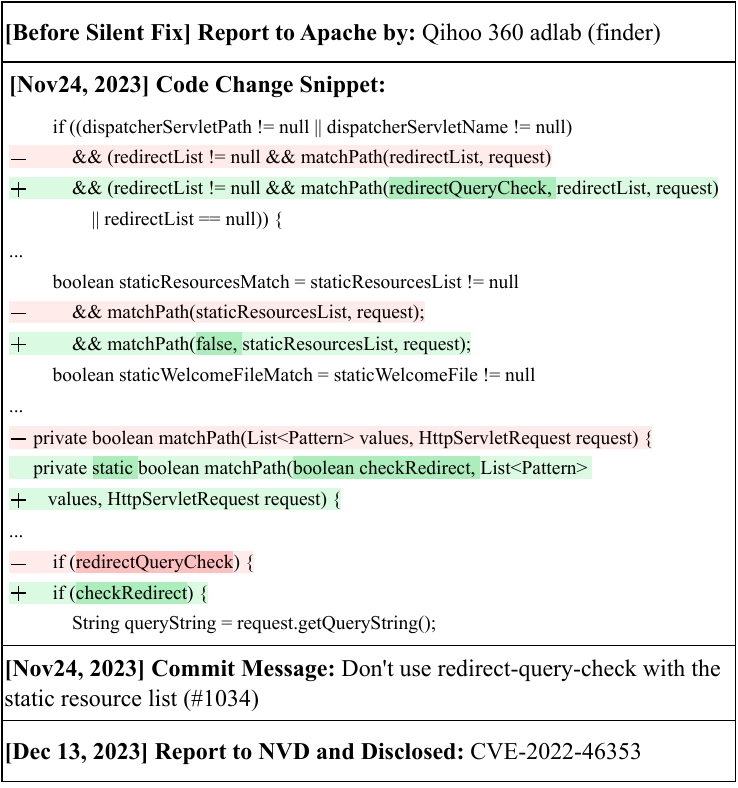}
	\caption{A Silent Vulnerability Fix for CVE-2022-46363~\cite{Apache2022CVE} detected by \ourTool.}
	\label{fig:examplerq4}
\end{figure}

For instance, \ourTool successfully detects a silent vulnerability fix commit while other baseline failed to identify. The commit was a silent vulnerability fix for CVE-2022-46363, as shown in Figure~\ref{fig:examplerq4}. The vulnerability was initially reported to Apache by Qihoo 360 adlab. Then, it was silently fixed without any indication of the vulnerability fix in the commit message on Nov 24, 2022, following the CVD procedure before it was revealed and published on NVD on Dec 13, 2022.

\rqboxc{\ourTool significantly outperforms the SOTA baselines \vfm, CoLeFunDa and \midas across all metrics in this temporal dataset setting. For instance, \ourToolStar improves the baselines at least by 64.5\% and 93.9\%, in terms of F1 and \costFive, respectively.}




\section{Discussion}\label{sec:dis}

\subsection{Potential applications of \ourTool on other tasks}
While our task is to identify silent vulnerability fixes, our framework is flexible. It can be used for other code change-related work, such as just-in-time defect prediction~\cite{hoang2019deepjit,hoang2020cc2vec,liu2023ccrep} and commit message generation~\cite{hoang2020cc2vec,liu2023ccrep}, which rely on capturing the code change information as the fundamental component. For instance, for the commit message generation, our code change representation could be used as the encoder and then connected to the decoder to generate the message. We encourage future research to investigate our code change representation to other tasks. 

\subsection{Real World implementation of \ourTool}
In RQ4, we evaluated \ourTool on the VFM2023 temporal dataset. Although \ourTool outperformed existing approaches, we observed that all methods, including ours, exhibited low precision. This reflects the inherent difficulty of the task in both industry and academia due to the highly imbalanced nature of the data. While manual code review is still necessary to verify the suggestions made by our tool, the development of such tools remains crucial in reducing the overall workload for security experts, who cannot realistically review every commit across all repositories. Considering more practical metrics like CostEffort, which reflect real-world usage scenarios, \ourTool demonstrates significant value. For instance, by reviewing only 5\% of the retrieved commits, security experts were able to identify 44\% of the vulnerability fixes. Feedback from our industry partners further supports the tool’s utility, as they found the positive rate to be acceptable, requiring only a modest amount of additional manual effort to review the results.

\subsection{Threats to Validity}

\noindent\textbf{Internal Validity}
In this study, we use two categories of metrics to evaluate the effectiveness of \ourTool and baselines. Although there might be other metrics that could be used, our selected metrics are commonly used in evaluation in previous studies~\cite {zhou2021vulfixminer,midas} for the same task. \ourTool has a parameter which is the size of the context window. We selected the window size empirically and investigated their impact in RQ3. In this study, we evaluate the impact of the size of context on one model with a max token limit of 1024. Although our experiment results suggest that using a context window of 3 lines provided the best result, and we observe a decline trend after 3 lines. However, more advanced models that support more tokens might perform better with a larger context window, and we suggest future research to investigate models with larger window sizes. Another threat is that we consider all files in a VF commit as VF when training file-level prediction model, this may not always be the case and introduce bias. To mitigate this, we used a widely used dataset VFM, and demonstrated the superiority of \ourTool over SOTA baselines.

\noindent\textbf{External Validity}
Threats to external validity relate to the generalizability of our findings. In this study, we evaluated our proposed framework on one task and demonstrated its superiority in identifying vulnerability fixes over baselines. Future research is encouraged to evaluate our approach to more code change-related tasks.

\section{Conclusion}\label{sec:conclusion}
In this paper, we propose a simple yet effective framework to identify commit-level vulnerability fixes by enhancing code change representation learning to improve the capacity of our model for capturing small code changes, and simplifying the training process to enable us to train the embedding models and final classification model jointly. Our evaluation shows that \ourTool outperforms SOTA baselines and achieves an F1 of 0.33, a \costFive of 0.632, and a \costTwe of 0.76, with an improvement of 77.4\%, 7.1\%, and 5.1\% over the best baselines, respectively. Our ablation analysis shows that our \codeDelta plays a critical role in \ourTool, and is typically effective in capturing small code changes. Evaluation of the temporal dataset demonstrates the superiority of \ourTool over baselines with a large margin improvement in real-world situation. 

\bibliographystyle{ACM-Reference-Format}
\bibliography{reference}

\end{document}